\documentstyle[prl,aps,twocolumn,epsf]{revtex}
\begin{document}

\title{Non conventional screening of the Coulomb interaction in low
dimensional and finite size system}

\author{J. van den Brink and G.A. Sawatzky}
\address{
Laboratory of Applied and Solid State Physics, Materials Science Center,\\
University of Groningen, Nijenborgh 4, 9747 AG Groningen, The Netherlands}
\date{\today}
\maketitle
\begin{abstract}
We study the screening of the Coulomb interaction in non polar systems
by polarizable atoms. We show that in low dimensions and small finite
size systems this screening deviates strongly from that conventionally
assumed. In fact in one dimension the short range interaction is
strongly screened and the long range interaction is anti-screened
thereby strongly reducing the gradient of the Coulomb interaction and
therefore the correlation effects. We argue that this effect explains
the success of mean field single particle theories for large
molecules.
\end{abstract}


In model Hamiltonians such as the Hubbard model set up to treat
correlation effects the on-site Coulomb interaction is a very important
parameter. It is usually argued that this interaction is only weakly
screened relative to the longer range interactions which  justifies
the neglect of the later~\cite{Hubbard63}. In order to decide on the importance of
the correlation effects precise knowledge of the longer range Coulomb
interactions is crucial since of course it is the gradient of this
interaction which really matters. In the last years for instance
correlation effects have been an important topic in the debate on high
T$_{\rm c}$ superconductors, colossal magneto resistance materials, Kondo and
heavy fermion systems, organic conductors and (doped) fullerenes.
Usually one uses dielectric response theory to estimate the range of
the Coulomb interaction and one comes to the conclusion that the
effective interaction is very short range without really asking
whether or not this theory is applicable. 
In this paper we study the
static dielectric response of a non polar insulator to the Coulomb
potential generated by two electrons and show that the above mentioned
approximation is very incorrect leading not only to quantitatively but
also qualitatively different physics.
We show that the well known Clausius-Mossotti (C-M)
theory of dielectrics which is valid for large distances in 3
dimensions breaks down at short distances and for lower dimensions in
general. At very short distances between two electrons the Coulomb
potential is a rapidly varying function of distance so that the local
field corrections have to be taken into account. We show that on one
and two dimensional systems (lines and planes of atoms) and large
molecules the screening of the (3 dimensional) Coulomb field is
totally due to local field corrections. For finite systems, like
molecules, the effective Coulomb interaction deviates strongly from
the $(\epsilon r)^{-1}$ like behavior assumed in dielectric response theory.
The dependence on distance turns out to be much weaker giving rise to
a more uniform potential for the electrons. In fact in low dimensions
the situation is quite the opposite to that usually assumed in that
the short range interactions are efficiently screened whilst the
longer range interactions are anti-screened. This results in a drastic
reduction of correlation effects, providing a rational for the success
of one electron mean field like theories to describe the electronic
structure of organic molecules.

We consider our systems to be composed of polarizable atoms.
If we assume a linear response of the atoms to an electric field, the induced
dipole moment ${\bf p}({\bf r}_i)$ on an atom on site ${\bf r}_i$ is
proportional to the local field ${\bf F}({\bf r}_i)$:
${\bf p}({\bf r}_i)=\alpha_i {\bf F}({\bf r}_i)$, 
where $\alpha$ is the atomic polarizability. 
Assuming the electric field is constant over the size of the atom, the energy 
of the system is lowered due to the induced dipole moment by an amount of 
$\Delta E_i = -\frac{1}{2} \alpha_i \ |{\bf F}({\bf r}_i)|^2$. Using this, 
the screening energy of the Coulomb
interaction of two electrons on the same lattice point (the Hubbard $U$)
can be obtained~\cite{Brink95_2}.

Clausius and Mossotti derived the relation between the microscopic
polarizability and the macroscopic dielectric constant
$\frac{\epsilon-1}{\epsilon+2} = 4 \pi  \alpha/3$,
assuming that the external potential
can be considered as being uniform in a region around 
each atom, and that one can treat this region as a continuum.
If the electrons are close to each other, one has to take into account 
that a lattice consists of a discrete set 
of atoms.
We show below that this has important consequences if the
distance between the electrons is of the
order of the lattice spacing or if the dimensionality of the system is low. 

The mathematical description of the microscopic response of a
system of point-dipoles to external fields was developed by Mott
and Littleton~\cite{Mott38} and has proven to be a powerful
method to calculate for
instance local field corrections for point-charge defects~\cite{Wang80} and
(surface) core-level shifts in 
photoemission~\cite{Boer84,Rotenberg}. 
Consider two electrons, one at
the origin of our coordinate system and one at ${\bf R}$. 
The aim is to calculate the electrostatic energy $V$ of the system as a function
of ${\bf R}$. The total energy consists of four parts: the potential
energy
due to the monopole-monopole, monopole-dipole and dipole-dipole interactions
plus the energy stored in the induced dipoles~\cite{Wang80}.
The total electrostatic energy is given by: 
\begin{equation}
V({\bf R}) = \frac{e^2}{R} - \frac{1}{2} \sum_{i} {\bf d}_i\cdot {\bf p}_i
           \equiv \frac{e^2}{R} - 2 E_p({\bf R}),
\label{eq:V(R)}
\end{equation}
where ${\bf d}_i$ is the total external monopole field and 
${\bf p}_i$ the induced dipole moment at site $i$. 
The bare Coulomb repulsion is reduced by the polarization
energy $2E_p({\bf R})$, as defined in the last part of 
equation~(\ref{eq:V(R)}). 
Now the dipole moment at each site has to be determined. 
The induced dipole moment
is proportional to the local field on the atom, leading to:
\begin{eqnarray}
{\bf p}({\bf r}) &=& \alpha e {\bf r}/|r|^3
                 + \alpha e ({\bf r}-{\bf R})/|r-R|^3 \nonumber \\
                 &+& \alpha \sum_{ {\bf l} , {\bf l} \neq 0 }
                   \frac{1}{|l|^5} \ \{ \ [3{\bf p}({\bf r}+{\bf l}) 
                   \cdot{\bf l}] \ {\bf l}
                 - |l|^2 {\bf p}({\bf r}+{\bf l}) \}.
\label{eq:p(r)}
\end{eqnarray}

This equality for $N$ atoms results in $DN$ coupled equations for 
${\bf p}({\bf r}_i)$, where $D$ is the dimensionality of the system.
If the symmetries of the problem are taken into account, the number of
equations that actually have to be solved can be substantially reduced.
The matrix representation of equation~(\ref{eq:p(r)}) in Cartesian coordinates 
is:
\begin{equation}
p_i^{\mu} = \alpha_i d_i^{\mu} + \sum_{\gamma} \sum_{j, j \neq i}
M_{i j}^{\mu\gamma} p_j^{\gamma}, \ {\rm with} \ \ \mu,\gamma = x,y,z.
\label{eq:p_matrix}
\end{equation}
The elements of the matrix that represents the dipole-dipole interactions
are given by:
\begin{equation}
M_{i j}^{\mu\gamma} = \alpha_i 
      ( 3 l_{ij}^{\mu} \ l_{ij}^{\gamma} \ |l_{ij}|^{-5}
       -|l_{ij}|^{-3} \ \delta_{\mu \gamma} ),
\label{eq:m_elem}
\end{equation}
where ${\bf l}_{ij}= {\bf l}_{i}-{\bf l}_{j}$ is the vector connecting the
two dipoles.  The solution of this set of 
equations gives the exact effective potential for the electrons. 
In some cases the equation can be solved in wave-vector space~\cite{Mahan80}, 
but in general the less cumbersome method is to solve the equation numerically 
in real space.

The formulation above reduces to the C.-M. relation when  a) the dipole-dipole
interaction is averaged over and  b) the lattice is treated as a continuum.
This is easily shown as from the C.-M. relation follows that
$
{\bf p}_{i} = \alpha_i \frac{\epsilon+2}{3\epsilon} {\bf d}_{i}
$
so that
$
2E^p ({\bf R}) =  \frac{\epsilon+2}{3 \epsilon} \frac{4 \pi \alpha e^2}{R}  
$
and the well know result 
$V({\bf R}) = {e^2}/{\epsilon R}$
is obtained if the summation over the lattice vectors is replaced by an integration. 
An obvious improvement up on the  C.-M. result is to evaluate the lattice sum in
equation~(\ref{eq:V(R)}) exactly.
Within this partial continuum limit of this model, where dipole-dipole interactions are taken
into account on the Clausius-Mossotti level,  values for the polarization 
energy and the effective on-site Coulomb interaction $U$ that compare well 
with experiment can be obtained~\cite{Brink_tbp}.

If one wants to take the dipole-dipole interaction into account without
using the Clausius-Mossotti relation, one has to solve the matrix equation
(\ref{eq:p_matrix}). 
Only a limited number of equations can be solved numerically, typically a few thousand.
For a 3D cubic system the exact effective potential deviates not more than around 10\% from the C.-M.
expression, except if the charges are at the same site, 
as has been discussed before~\cite{Brink95_2}.

\begin{figure}
      \epsfysize=60mm
      \centerline{\epsffile{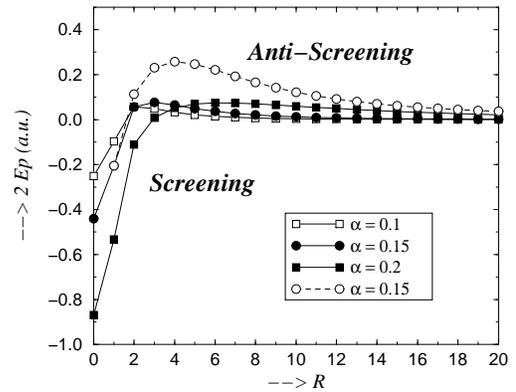}}
\caption{The polarization energy $2 E_p$ as a function of separation between
the electrons in one dimension.  The lattice spacing is taken
to be unity. The  dashed line represents $2 R E_p$ for $\alpha=0.15$.
Negative values of $E_p$ represent screening and positive values anti-screening.
}
\label{fig:epol1D}
\end{figure}

The situation is very different for a system where the dipoles and electrons are confined to a plane 
(2D system) or a line (1D system), a  geometry that can be encountered in a material where a chain 
of polarizable  atoms is embedded in a  non polarizing matrix.
As the system is an object in the 3D real space, the bare Coulomb
interaction between the two electrons is inversely proportional to the distance between the
charges. 
For a 1D system the polarization energy is calculated exactly and shown 
in figure~\ref{fig:epol1D}.
At very short distances (till about 2 lattice spacings) the Coulomb interaction 
is screened.
When the separation between the charges is larger, however, 
the Coulomb interaction is {\it anti-screened}: the induced polarization
results in an increased repulsion between the two charges. This behavior is
markedly different from the C.-M. result.
In fact it is the opposite to the usually assumed behavior that long
range interactions are screened and short range interactions not or
weakly.

\begin{figure}
      \epsfysize=40mm
      \centerline{\epsffile{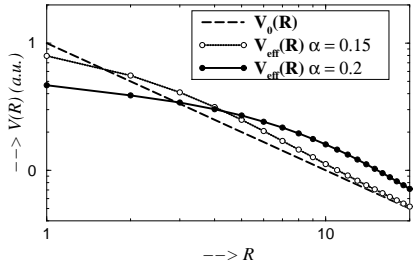}}
\caption{Effective Coulomb potential in a 1D system with
$\alpha=0$, $\alpha=0.15$ and $\alpha=0.2$ on a Log-Log scale. The dashed straight
line represents the unscreened Coulomb interaction. Screening tends to flatten the
effective potential at intermediate distances. 
The lattice spacing is taken to be unity.}
\label{fig:v_eff1D}
\end{figure}

In the figure also $R E_{\rm p}$ as a function of distance is shown. 
If a dielectric constant $\epsilon$ can be defined 
then $2 R E_p = 1 - 1/ \epsilon$. From the figure
it can be seen that $2 R E_p \rightarrow 0$
when $R \rightarrow \infty$, so that at large distances  
$\epsilon \rightarrow 1$ and the 
Coulomb potential is unscreened. As a consequence the effective
potential between the charges is flattened out. At short distances the bare
Coulomb potential is large, but also the polarization energy is large, see
figure~\ref{fig:v_eff1D}. 
At intermediate distances the Coulomb potential decays, but the polarization
enhances the repulsive interaction so that in the end the effective potential decays
much slower than 1/R. 
In a two dimensional system an analogue situation is encountered. At short
distances local fields screen the potential and at large distances the potential is 
again unscreened.

Why is this result so different from what one expects on the basis of the C.-M. relation?
If the distance between the electrons is large the lattice can be treated as a continuum.
and the lattice sum can be replaced by an integral.
In the low dimensional systems the outcome of the integral does not only 
depend on the distance between the two electrons,
but also on the range of integration. In order to make the integral convergent
we have to assume a sphere around each electron with radius $r_0$, in 
which the polarizability is zero. 

\begin{figure}
      \epsfysize=40mm
      \centerline{\epsffile{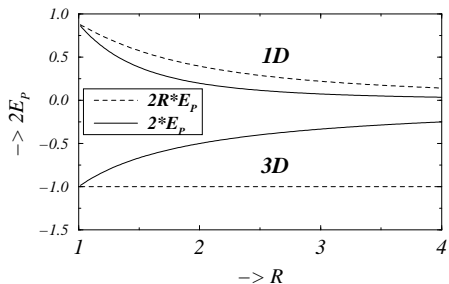}}
\caption{The polarization energy $2 E_p$ as a function of separation between
the electrons in one and three dimensions in the continuum limit.  
The lattice spacing is taken to be unity and in the 1D system $r_0 = 0.5$.}
\label{fig:screen1_3D}
\end{figure}

In figure~\ref{fig:screen1_3D} the outcome of the continuum integral as a function of separation between
the electrons is plotted. For clarity the proportionality constants are set to unity. 
The behavior for the 3D system is as expected: the 
polarization energy is negative and proportional to $1/|R|$. In the 1D
system, however, the polarization energy is positive, implying that screening
effects {\it increase} the repulsion between the charges. This explains the
anti-screening of the Coulomb repulsion at large distances found in the exact calculation.

The figure also shows that the polarization energy decays faster than $1/|R|$ at large distances.
This can be expected since in 1 and 2D the integration is not over a volume, but rather 
over a line and a surface respectively. 
The consequence is that in low dimensional systems the Coulomb potential is unscreened at
large distances.

The exact calculation for the 1D system shows that the screening energy is negative at short distances.
This can be understood by considering the limiting case where the two charges are on the same
atom. The polarization energy associated with a single electron is a constant times the square of
its electrical field $-cF^2$, so that the total polarization energy of a system where two
electrons are infinitely far apart is $-2cF^2$.
If the two electrons are on the same site, the total field is $2F$ resulting in a polarization 
energy of $-4cF^2$.
So the screening energy related two charges at the
same site is always more negative than the screening energy of two charges far apart, explaining
why at short distances the Coulomb potential is always screened.

\begin{figure}
      \epsfysize=60mm
      \centerline{\epsffile{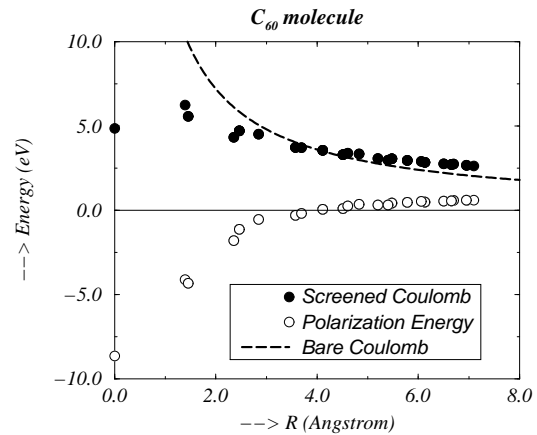}}
\caption{The effective Coulomb interaction as function of the separation of two electrons
on a C$_{\rm 60}$ molecule. The carbon-carbon distance is 1.391 \& 1.455 \AA \ and the carbon
polarizability is taken to be 0.56 \AA$^3$}
\label{fig:C60}
\end{figure}

We now turn to a few finite size systems.
For a C$_{\rm 60}$ molecule
we can model the screening by assuming that on each carbon site a dipole moment can be
induced due to two excess charges on the molecule. The effective Coulomb repulsion between
these two charges is calculated by the method described above.
The results for the C$_{\rm 60}$ molecule is shown in figure~\ref{fig:C60}. Whereas the bare Coulomb repulsion
depends strongly on the distance between the charges (ranging from $\sim$13.5 to $\sim$2 eV), screening tends to
flatten the effective interaction (ranging from $\sim$6 to $\sim$3 eV). This is again due to the fact that at
short distances the Coulomb interaction is screened and at large distances anti-screened, as in the
1D case. This behavior compares well with what has been found by Gunnarsson {\it et al}~\cite{Gunnar92}
and Lof {\it et al}~\cite{Lof95}.

\begin{figure}
      \epsfysize=60mm
     \centerline{\epsffile{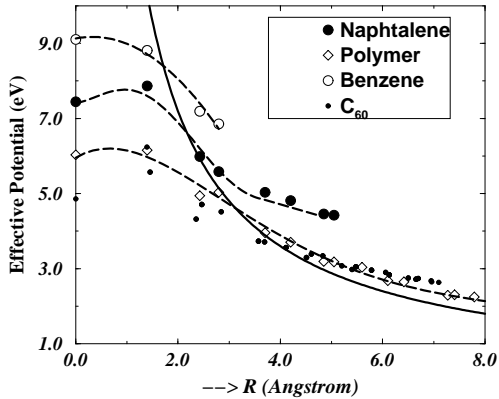}}
\caption{The effective Coulomb interaction on different organic molecules. 
The carbon polarizability is  0.56 \AA$^3$. The full line represents the
bare Coulomb repulsion. The dashed lines are guides for the eye.}
\label{fig:molecules}
\end{figure}

In figure~\ref{fig:molecules} the results for benzene and two linear benzologues, naphthalene and
an polymer consisting of 200 benzene rings, is shown. The carbon-carbon distances and carbon polarizability are
taken from reference~\cite{Silinsh80}. The Coulomb potential for larger molecules
is screened more effectively because more atoms participate in the screening process.
For all these systems, however, the variation of the effective Coulomb potential over the molecule is
drastically reduced by screening effects.
For distant neighbors there is anti-screening, since the introduction of an electron
on the molecules moves screening charge to the opposite site of the molecule causing the
screening clouds of the two charges to interfere destructively. Obviously this makes the C$_{\rm 60}$
molecule and the linear benzologues less correlated: the electrons move in a relatively uniform 
effective potential. Interesting also is that the on-site and nearest neighbor Coulomb interactions
in these systems are almost equal, which really makes a Hubbard-like
description very questionable.
This explains why one-particle theories that do not take correlation effects fully into account, 
work so well for large organic molecules.

We considered, in conclusion, a point-dipole model to account for the screening of the Coulomb 
repulsion in non-polar insulators.
For three dimensional systems the deviations from the Clausius-Mossotti result
for the inter-site screened Coulomb interaction are small, but
in one and two dimensional systems only local field effects contribute to the
screening. At large distances the Coulomb interaction is unscreened and in 1D at intermediate
distances the Coulomb interaction is even anti-screened.
Applying the dipole screening model to finite size systems like large organic molecules, we showed that the
effective Coulomb interaction is only weakly dependent on the distance between the electrons.
So correlation effects are drastically reduced, explaining the success of, one-particle, molecular
orbital theory for large molecules. 

This work was financially supported by the Nederlandse Stichting voor Fundamenteel Onderzoek der Materie (FOM)
and the Stichting Scheikundig Onderzoek Nederland (SON), both financially supported by the Nederlandse
Organisatie voor Wetenschappelijk Onderzoek (NWO).

\end{document}